# Leveraging LLMs for Design Ideation: An AI Tool to Assist Creativity

Rutvik Kokate[1][0009-0000-9998-6110] , Pranati Kompella[1][0009-0004-3191-0100] , and Prasad Onkar[1][0000-0003-1788-3826]

[1]Indian Institute of Technology Hyderabad, Hyderabad, India
rutvikokate2610@gmail.com
pranati.kompella@gmail.com
psonkar@des.iith.ac.in

**Abstract.** The creative potential of computers has intrigued researchers for decades. Since the emergence of Generative AI (Gen AI), computer creativity has found many new dimensions and applications. As Gen AI permeates mainstream discourse and usage, researchers are delving into how it can improve and complement what humans do. Creative potential is a highly relevant notion to design practice and research, especially in the initial stages of ideation and conceptualisation. There is scope to improve creative potential in these stages, especially using machine intelligence. We propose a structured ideation session involving inspirational stimuli and utilise Gen AI in delivering this structure to designers through ALIA: Analogical LLM Ideation Agent, a tool for small-group ideation scenarios. The tool is developed by enabling speech based interactions with a Large Language Model (LLM) for inference generation. Inspiration is drawn from the synectic ideation method and the dialectics philosophy to design the optimal stimuli in group ideation. The tool is tested in design ideation sessions to compare the output of the AI-assisted ideation sessions to that of tradi tional ideation sessions. Preliminary findings showcase that participants have rated their ideas better when assisted by ALIA and respond favourably to speech-based interactions.

## 1. Introduction

Good ideas are the basis for creative design processes. Pioneers of studies in design creativity have defined creativity in terms of the capacity to produce novel, unique, and useful ideas [1–3]. Consequently, the pursuit of enhancing design creativity often employs methods that promote or enhance idea generation. There are several ideation methods and techniques designers employ to generate ideas as part of the design process [4, 5]. Research shows that professionals and experts use a variety of cognitive structures to assist their ideation [6]. Heuristics are one such cognitive structure [7, 8]. Internal and external stimuli are also often used as sources of inspiration. Internal stimuli can take the form of mental imagery or verbal information [9]. Text as a representational stimulus has shown a positive influence during idea generation when compared to an environment with no external stimuli [10]. The stimulus for inspiration often depends on the idea-generation method chosen. The objective of this study is to support design ideation sessions using inspirational stimuli. We conceptualise and propose an ontology for a design ideation session using inspirational stimuli based on cognitive heuristics. The study derives heuristics from Gordon's synectics [11] and Hegel's dialectic [12]. The intent is for designers to be able to employ this ontology in their ideation process. The rise of Gen AI has augmented human creativity, particularly in situations where inspiration and idea generation are in demand. This collaboration between humans and AI agents is gaining traction in the domain of computational co-creativity. Research has examined the role and utility of generative AI in design ideation sessions [13] as supportive technology to augment creativity and has revealed several insightful attributes of AI-assisted idea generation along with some limitations as well. To understand what AI can do for design ideation, we prototyped the application ALIA: Analogical LLM Ideation Agent. ALIA augments human creativity during ideation by providing timely inspiration for participants in an ideation session. Three pilot studies were conducted to primarily test the functionality of the tool and validate the effect of its intervention on the participants. We plan to evaluate ALIA via experiments on two fronts: to observe participants' reactions to ALIA, particularly how they perceive the inspirational stimuli provided and how they assess their own ideas after ideation. Secondly, to analyse and compare the ideas produced with and without ALIA. This paper covers the first objective of our evaluations, i.e. participants' reactions. In this work, all the products of an ideation session are ideas, whether or not they directly address the given problem statement. Thus, in experiments and evaluations, participants produce leads, concepts, and various notions as "ideas," along with concrete concepts or solutions. This paper is structured as follows: Sect. 1 provides an introduction, Sect. 2 gives the research background for our work, Sect. 3 details the methodology followed, Sect. 4 reports the results of experiments, Sect. 5 details the discussion and future work, Sect. 6 gives the conclusion of the study.

## 2. Literature review

Literature review for this study situates the research topic within existing studies of how design creativity has evolved and its interaction with cutting-edge technologies like Gen AI and computer-aided creativity. State-of-the-art studies on Gen AI for design ideation reveal an emerging field of human-AI collaboration for ideation and the gaps provide critical insights into how it can be effectively utilised in creative ideation context. The role of a structured creative process and its contribution towards an effective ideation session is reviewed with the study of existing philosophical frameworks for ideation.

### 2.1 Design Creativity and Computer-Aided Creativity

The notion of creativity is highly relevant in a wide spectrum of fields like engineering, entrepreneurship, management, and research. Product development, innovation, and ideation engage human creativity. Due to pressure on people participating in product development to churn out ideas quickly, many similar and redundant ideas are produced and launched into the market [14]. Creativity plays the most critical role in determining the success of an idea and distinguishing it from others. Researchers have grappled with creativity for decades [15, 16]. The computational modelling of convergent and divergent thinking allows the demystification of creative cognition [17, 18]. In this work, we deal with creativity in design ideation. Researchers have similarly divided conceptual design into two stages: divergent and convergent concept generation [19, 20]. Divergent implies high abstraction, while convergent implies low abstraction. Liu et al. proposed an "ideal" method for concept generation that involved a process of repeated convergence and diver gence [17]. As computational creativity is reaching new bounds, computer-aided creative systems are being established. [3] proposed a computer-aided creative design platform with a web search tool, a knowledge base, a conceptual design space, and creative methods. They do note in their article that computer-based systems have various limitations, so human-based creativity seems to be the most pragmatic approach. [18] categorised the different ways in which computers support design tasks. They first delineate the generation, evaluation, and selection of design solutions. They subsequently categorise the resulting interactions between computer and designer as computer-aided (providing feedback or facilitating tasks), computer-based (automation), and computer-augmented (extending the designer's capabilities). In this work, we examine the generation stage in design ideation and facilitate computer-augmented interactions with our tool.

## 2.2 AI in Design Ideation

Since the emergence of Gen AI, the creative capacity of computers has taken on many new applications and has significant potential to augment human tasks. Tholander et al. [13] observed that integrating AI in design ideation has been helpful in speeding up the design process. However, the participants found the AI responses meaningful in interpreting the design task rather than being a good design intervention. The participants reported not trusting the AI to generate high-quality solutions on its own. They perceived the AI system as useful for thinking broadly rather than going deep into a design space. Such AI systems generally lack contextual reference to participants' intentions and rationales. Memmert and Tavanapour [21] studied human-AI collaboration in general brainstorming scenarios and found significant potential for cognitive stimulation in humans by interacting with the AI, but also a risk of free riding solely on the AI responses. Lavric and Skraba [22] proposed a technical framework for human-AI integration to produce innovative ideas. They highlight some concerns with AI-generated ideas; some may be ethically questionable, and the explainability of generated responses is challenging. They also highlight the relevance of the temperature parameter for the innovativeness of ideas despite the ambiguity in its specifications. The use of GPT-3 in design ideation and how it both limits and enables human–machine co-creativity. Humans interacted with the LLM via text (typing) in their study. In such a text-based interaction, humans break their ideation activity to perform the physical activity of typing. They also spend time and mental resources thinking about the right words to input to the LLM. We argue that these activities could centre the ideation too much around the LLM responses rather than the human discussions. This research gap is potentially significant. Thus, in our work, we enabled speech and display-based interactions between humans and the ideation tool in our project not unnecessarily to disrupt the humans' ideation.

## 2.3 Philosophical Frameworks: Synectics and Dialectics

Synectics [11] is an ideation methodology proposed by Gordon, W. J. where the core principle is to embrace the seemingly irrelevant, which is achieved by alienating the original problem context through analogies. Going down analogous trains of thought and abstracting the context can provide unexpected inspiration. It promotes divergent thinking [20] to pick up tangential, random, loosely related lines of thought and move away from the obvious problem context. It outlines some heuristics that one can use in different stages of the ideation session, characterised as "idea triggers" [7]—"add, subtract, transfer, superimpose, fragmentate, and empathise". Some are very specific and concrete, while others are more general. Dialectics is used to describe a philosophical argument that involves a contradictory process between opposing sides [23]. Hegel's dialectics [12]—a process of thesis, antithesis, and synthesis is used to describe a philosophical argument that involves contradictory ideas forming tension and resolution, which propels the idea towards an innovative solution. These inconsistencies provide compass points for examining and balancing conflicting elements within the context of the issue, facilitating the identification of viable solutions. In this work, dialectics are used as heuristic instruments for ideation.

## 3. Methodology

This work proposes a structured design ideation session involving inspirational stimuli based on certain cognitive heuristics. An essential component of the structure of this ideation session is repeated convergence and divergence, as discussed in Sect. 2.1. One of the goals was to nudge participants, through stimuli, to engage in divergent thinking, then narrow their focus, diverge again, and continue this process. Based on the philosophical frameworks, an ideation structure was formulated.

### 3.1 Stages of Ideation

The structure of the ideation session is designed such that one session is roughly divided into four stages. A breakdown of the ideation session into its stages is shown in Fig. 1. In the first stage of the session, we introduce the problem statement to the participants and display a couple of starter questions. In the second stage, the participants are taken on an "excursion" (a notion derived from synectics, discussed in Sect. 2.3) by providing them with a list of words loosely related to their recent discussion. The participants look over the words and decide upon a word that will guide their subsequent conversation. In the third stage, analogical questions (derived from synectics) are used to nudge their discussion. These questions are based on the discussion participants have in stage two. In the fourth stage, prompts are based on dialectics (discussed in Sect. 2.3); for participants to synthesise conflicting sides of the problem with their ideas. In this stage, participants are presented with a thesis and an antithesis (framed as questions) and are urged to arrive at syntheses. A desktop prototype was built that offers structured, contextually aware stimuli suited to certain situations to assist participants in ideation sessions. The philosophical frameworks mentioned in Sect. 2.3 were used as a foundation to provide inspirational stimuli. Based on the opportunities and role of LLMs in ideation activities, the prototype utilises a pre-trained LLM for generating stimuli in the form of textual responses. The following subsection describes the process of prototype development (see Fig. 1).

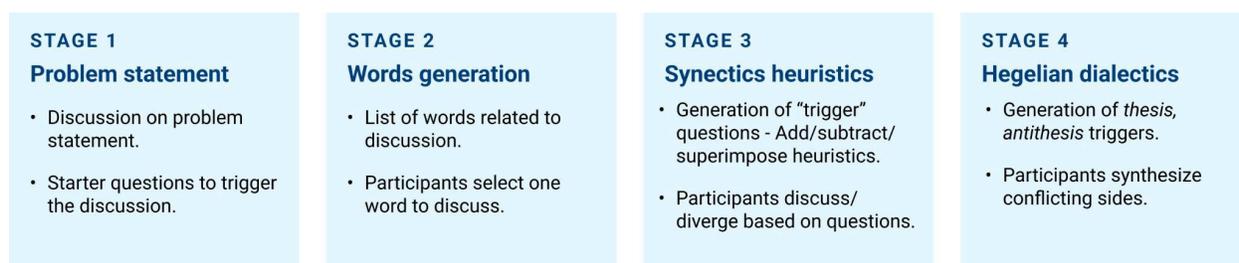

**Fig. 1.** Description of ideation stages.

### 3.2 Prototype Development

The prototype—ALIA consists of a digital application and hardware sensors and is the primary component with which the users interact. ALIA's interface comprises four screens, each representing one of the stages (see Fig. 1) that display stimuli during ideation. Given the limitations in typing-based interaction with LLMs (discussed in Sect. 2.2), the prototype eliminates the need for text-input prompts to the LLM from the people engaged in ideation. ALIA uses a speech-to-text transcriber to generate transcriptions of discussions in real time, and a summariser generates periodic summaries of the ongoing discussion. Based on the philosophical frameworks and ideation stages (Sect. 3.1), underlying LLM is provided with contextual understanding (summary) of the participant discussion as prompts (detailed description in Sect. 3.4) for inference generation. Participants can hover their hand over an LDR (Light dependant resistor) placed nearby, which act as a mode of input to control response generation and switching between the screens.

## 3.3 Pilot Study Findings

After the first conceptualisation of a guided ideation session, an initial prototype was built. Three pilot studies were conducted to test the functionality of this application. The "idea triggers" described in Sect. 2.3 were tested in these studies. Based on the feedback, the following triggers were found to be useful during the session: Subtract: involves reducing the problem or removing parts; Add: involves extending or expanding the problem to consider more; Superimpose: involves looking at the problem from a new perspective. Findings reported participants being abruptly breaking their train of thought whenever a screen changed to next stage without prior intimation. Participants frequently paused to reflect, but they were less disruptive during conversational breaks. With this feedback, two conditions were set to be met for the system to display a nudge: first, the minimum allocated time for each stage has elapsed, and second, a minimum silence of 8 s is detected. Once participants gave their consent to move forward (by touching an LDR placed nearby), the system moved forward

## 3.4 Prompts

A critical component of ALIA is the structure in which the prompts are designed.. Based on the work outlined in [24] a prompting framework—COSTAR (Context, Objective, Style, Tone, Audience, Response format) is utilised. COSTAR prompting is applied in ALIA to structure the content and set contextual information to the model. Prompt designing was an iterative process. COSTAR prompting is applied in ALIA to structure the content and provide information about the structure described in Sect. 3.1 to the model. All iterations of prompts at each stage were documented. The prompts were tweaked to get the desired response. An example of COSTAR prompt used in the system is shown in Fig. 2.

```
# CONTEXT #
There is a brainstorming session going on. Synectic method of brainstorming by JJ Gordon is being used in the session. The goal of this idea
generation process is to help the participants explore, discuss different perspectives, backgrounds to generate a wide range of creative ideas
in novel and effective ways. You are a part of this acitivity and you have to keep the session engaged and interesting for the participants.
#########
# OBJECTIVE #
Generate a list of 8 single-worded words based on the context of participant's conversation summary : {summary}.
\n The words should be short and concise. Generate only the words and nothing else.
#########
# STYLE #
Write in an informative and instructional style. Keep the words short and easy to read quickly for the participants.
#########
# TONE #
Maintain a positive and motivational tone throughout, fostering a sense of empowerment and encouragement.
# AUDIENCE #
The target audience is bachelors first year students interested in having idea generation discussions on the problem statement :
{problem_statement}.
\n There are two participants so use simple language to keep the information relevant.
#########
# RESPONSE FORMAT #
Provide a list 8 single-worded words. Provide only the words and no description.
```

**Fig. 2.** Example of COSTAR prompt

## 3.5 Experiment Setup

A set of experiments was conducted with two objectives: first, to see how people reacted to ALIA and second, to see how they perceived their performance in the ideation session with its presence. The experiments were divided into two categories—the control group, i.e. without the intervention of ALIA, and the experimental group, i.e. using ALIA in ideation session. Both categories involved 6 different groups, where each group had 2 participants, with the exception of one control group, which had 3 participants. Each of the six groups were presented with different problem statements for the idea generation activity. The same six problem statements were repeated for control and experimental groups, where each session was approximately 30 mins duration. Participants were university students between the ages of 18 and 20, somewhat familiar with brainstorming and ideation techniques. No explicit instructions were given to the control group participants regarding how to go about the ideation, and they were allowed access to internet resources. The experimental group was instructed about ALIA's usage and its role as an assistive system. The participants were reminded that the ideation was centred around them and were asked to disregard or ignore any stimuli produced by ALIA that confused or did not resonate with them and focus on their ideation discussion. The experiments were conducted in a closed, soundproofed room. Participants were provided with the ALIA system and ta able for placing sticky notes. Video and audio of all experiments were recorded. Participants filled out a feedback form post-experiment and gave detailed verbal feedback. Figure 3 shows the experiment setup along with the participants.

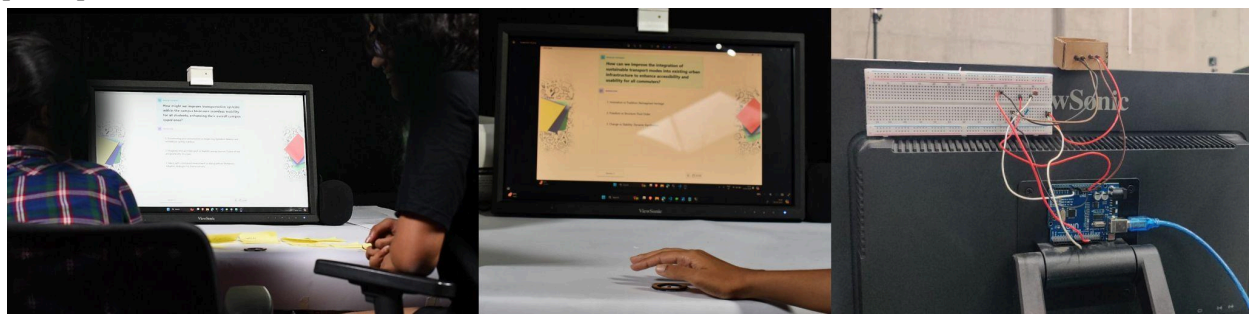

**Fig. 3.** Participants in an ideation session and hardware setup of ALIA

## 4. Results

The preliminary findings reported in this section come from the feedback collected from participants after each experiment. A few essential insights regarding each stage of the session stood out from this input. In Stage 1, participants found leading questions helpful as conversation starters. Participants could understand the given problem statement. In Stage 2, where ten keywords related to their discussion were displayed, participants took some time to finalise the word and continued with the discussion. In Stage 3, participants could adapt their discussion and diverge using the suggested triggers. Some groups found the questions too broad and conceptually distant from the problem statement and struggled to relate. Some groups took up the conceptually distant triggers to lead their discussion in a very different direction. The contradictions generated in Stage 4 helped the participants synthesise the points they discussed. Figure 4 displays how participants rated their ideas and their engagement with ALIA. A greater percentage of participants rated their ideas as "good" in the experimental group than in the control group. About 90% of the participants were engaged in "medium" and "high" levels of interaction with ALIA. Overall, the feedback from preliminary findings suggests the tool provides an adequate level of assistance and augmentation of thinking that the participants may need during their ideation session (Fig. 4).

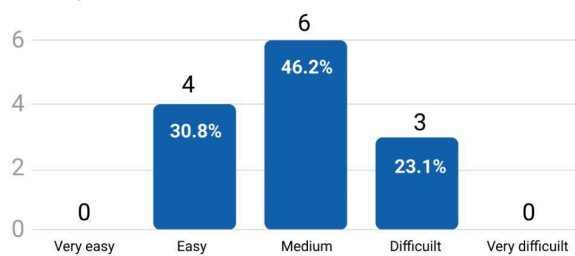
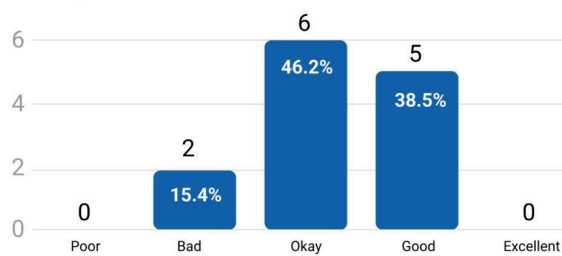
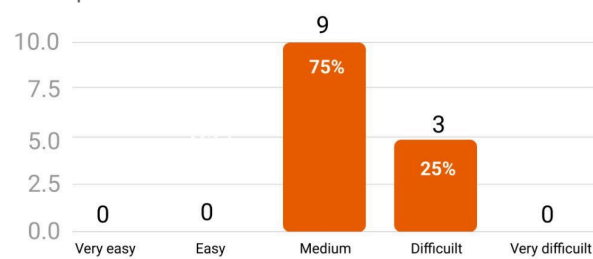
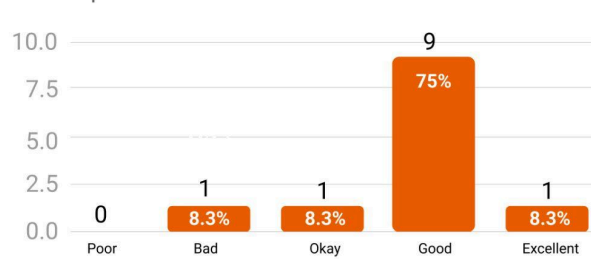

**Fig. 4.** Participants' feedback on ideation sessions

## 5. Discussion and Future Work

There are several limitations to this study—first is the demographics of the participant pool. Fresher design undergraduates were chosen to examine if AI could guide them with techniques that experts may already be familiar with. However, testing the tool with participants across different domains and age groups could lead to more nuanced insights on the utility of this tool in professional settings. The second is the choice of the LLM model utilised—Mistral 7b; with adequate financial resources, future testing on other LLM models could determine whether the tool is modular and applicable across other powerful models. A comprehensive evaluation of the tool must be done on two fronts: first, an examination of the ideas produced by using the tool vis-à-vis a standard set of ideas (produced during control experiments), and second, a deeper analysis of the quality of interaction between the participants and the tool. Further steps include conducting a thorough analysis of the ideas for their novelty and the pattern of convergence-divergence phases. Studying these patterns can offer insights into participant behaviour and the approaches participants take to creative thinking in an AI-assisted environment. Identifying distinct attributes of the AI-assisted set of ideas can provide insights towards the precise role and contexts in which LLMs might aid creative ideation.

# 6. Conclusion

In this work, we propose a structured ideation session that uses inspirational stimuli based on cognitive heuristics. Generative AI is posed as an aide that can assist humans by generating these inspirational stimuli through the application ALIA: Analogical LLM Ideation Agent. The AI system provides timely inspiration, unexpected leads and ideas during ideation and assists the participants in following the proposed structure. Additionally, condition-based speech interaction proved to be less interruptive during ideation sessions. Though LLMs may not be suitable for novel and highly creative ideas when utilized independently, they can be leveraged for divergent ideas in a human-collaboration setting. From the experiments, participants' feedback and reflections reveal that ALIA-assisted ideas are rated higher and found the inspirational stimuli to be helpful in their ideation. A comprehensive analysis of the verbal feedback collected is required to shed more light on the interaction with the AI system. Further, an analysis of the ideas produced with and without AI assistance is required to draw conclusions regarding the contribution of the proposed structure and the role of Gen AI in augmenting human creativity. Ultimately, this work aids the ongoing efforts in harnessing Gen AI to enhance human creativity and specifically investigate the creative potential of LLMs. The results of this study would be useful in the potential utilisation of AI assistance in ideation sessions.